\documentstyle{article}
\begin{document}
\newcommand{\beq}{\begin{equation}}
\newcommand{\eeq}{\end{equation}}
\newcommand{\beqn}{\begin{eqnarray}}
\newcommand{\eeqn}{\end{eqnarray}}
\newcommand{\dpf}{\displaystyle\frac}
\newcommand{\no}{\nonumber}
\newcommand{\ep}{\epsilon}
\begin{center}
{\Large A possible explanation of the clash for black hole entropy in the extremal limit }
\end{center}
\vspace{1ex}
\centerline{\large Ru-keng Su$^{1,2}$,\ Bin Wang$^{1,2}$,\ P.K.N.Yu$^3$ and E.C.M.Young$^3$}
\begin{center}
{$^1$ China Center of Advanced Science and Technology (World Laboratory),
P.O.Box 8730, Beijing 100080, P.R.China\\
$^2$ Department of Physics, Fudan University, Shanghai 200433, P.R.China\\
$^3$ Department of Physics and Materials Science, City University of Hong Kong,
Hong Kong}
\end{center}
\vspace{6ex}
\begin{abstract}
It is shown that the classical entropy of the extremal black hole depends on two different limits
procedures. If we first take the extremal limit and then the boundary limit,
the entropy is zero; if we do it the other way round, we get the Bekenstein-Hawking
entropy. By means of the brick wall model, the quantum entropy of scalar field in the extremal black
hole background has been calculated for the above two different limits procedures.
A possible explanation which considers the quantum effect for the clash of black hole
entropy in the extremal limit is given.
\end{abstract}
\vspace{6ex}
\hspace*{0mm} PACS number(s): 04.70.Dy, 04.62.+v
\vfill
\newpage
There has been a lot of papers discussing the entropy of the black hole in the
extremal limit recently, but the results are different[1-8]. Attention has been focused on whether the formula
of the Bekenstein-Hawking entropy is still valid or the entropy is zero for the
extremal black hole (EBH). Since this clash affects not only the statistical
interpretation of entropy but also the phase transition[9-12] of the black hole in the extremal limit,
it is of interest to further discuss this problem and investigate why contradictory results
have been obtained.

To illustrate the reason behind the contradiction, we study the two dimensional
(2D) charged dilaton black hole(CDBH)[7,13-15]. The action is
\beq                  
I=-\int_M \sqrt{g}e^{-2\phi}[R+4(\bigtriangledown\phi)^2+\lambda^2-\dpf{1}{2}
  F^2]-2\int_{\partial M}e^{-2\phi}K
\eeq
has a black hole solution metric
\beqn         
{\rm d}s^2=-g(r){\rm d}t^2+g^{-1}(r){\rm d}r^2\\
g(r)=1-2me^{-\lambda r}+q^2e^{-2\lambda r}      \\
e^{-2\phi}=e^{-2\phi_0}e^{\lambda r},\ A_0=\sqrt{2}qe^{-\lambda r}
\eeqn
where $m$ and $q$ are the mass and electric charge of the black hole respectively.
The horizons are located at $r_{\pm}=(1/\lambda)\ln(m\pm \sqrt{m^2-q^2})$.

Using the finite-space formulation of black hole thermodynamics, employing the
grand canonical emsemble and putting the black hole into a cavity as usual[3,7,16],
we calculate the free energy and entropy of the CDBH. To simplify our calculations, we
introduce a coordinate transformation
\beq    
r=\dpf{1}{\lambda}\ln[m+\dpf{1}{2}e^{\lambda(\rho+\rho_0^*)}+\dpf{m^2-q^2}{2}
    e^{-\lambda(\rho+\rho_0^*)}]
\eeq
where $\rho_0^*$ is an integral constant, and rewrite Eq.(2) to a particular gauge
\beq     
{\rm d}s^2=-g_{00}(\rho){\rm d}t^2+{\rm d}\rho^2
\eeq
Transformation in Eq.(5) can be used for both the non-extremal black hole(NEBH) and the EBH, because
in the extremal limit, the horizon $\rho_+$ and $\rho_-$ become degenerate and are
equal to $\rho_+=(1/\lambda)\ln\sqrt{m^2-q^2}-\rho_0^*$, so it changes to negative
infinity and satisfies the topological requirement of the EBH. The Euclidean action
takes the form
\beq         
I=-\int_{\partial M} \sqrt{\dpf{1}{g_{11}}}e^{-2\phi}(\dpf{1}{2}\dpf{\partial _1 g_{00}}
   {g_{00}}-2\partial_1 \phi)
\eeq
The dilaton charge is found to be
\beqn          
D=e^{-2\phi_0}(m+\dpf{1}{2}e^x+\dpf{m^2-q^2}{2}e^{-x})\\
x=\lambda(\rho+\rho_0^*)
\eeqn
The free energy, $F=I/{\beta}$, where $\beta$ is the proper periodicity of
Euclideanized time at a fixed value of the special coordinate and has the form
$\beta=1/T_w=\sqrt{g_{00}}/T_c$ is the inverse periodicity of the Euclidean time
at the horizon
\beq         
T_c=\dpf{\lambda\sqrt{m^2-q^2}}{2\pi(m+\sqrt{m^2-q^2})}
\eeq
Using the formula of entropy $S=-(\partial F/\partial T_w)_D$, we obtain
\beq             
S=\dpf{2\pi e^{-2\phi}[m+\dpf{e^x}{2}+\dpf{(m^2-q^2)e^{-x}}{2}][1+(m^2-q^2)e^{-2x}]
   \sqrt{m^2-q^2}(m+\sqrt{m^2-q^2})}{(m^2-q^2)+m[\dpf{e^x}{2}+\dpf{(m^2-q^2)e^{-x}}{2}]}
   \eeq
   Taking the boundary limit $x\rightarrow x_+=\lambda(\rho_+ +\rho_0^*)=\ln\sqrt{m^2-q^2}$
   in Eq.(11) to get the entropy of the hole, we find
   \beq         
   S=4\pi e^{-2\phi_0}(m+\sqrt{m^2-q^2})
   \eeq
   This is just the result given by Nappi and Pasquinucci[14] for the non-extremal CDBH,
   which confirms that our treatment above is right for the entropy of the black hole.

   We are now in a position to entend the above calculations to EBH. We are facing
   two limits, namely, the boundary limit $x\rightarrow x_+$ and the extremal limit
   $q\rightarrow m$. we can take the limits in different orders: (A) by first taking
   the boundary limit $x\rightarrow x_+$, and then the extremal limit $q\rightarrow m$;
   and (B) by first taking the extremal limit $q\rightarrow m$ and then the boundary
   limit $x\rightarrow x_+$. Obviously, taking the limits in these two different
   orders corresponds to two different treatments.
   In case (A), we first put a NEBH in a cavity and
   calculate its entropy, and then by taking the extremal condition to make the NEBH
   become EBH. In this treatment, we inherit the non-extremal topology of the black hole
   until to take the extremal limit.
   In case (B), we put an EBH into a cavity and initiate our work
   with the extremal topology. This treatment is similar to that of refs.[1,2].
   To do our limits procedures mathematically,
   we may take $x=x_+ +\ep, \ep\rightarrow 0^+$ and $m=q+\eta, \eta\rightarrow 0^+$,
   where $\ep$ and $\eta$ are infinitesimal quantities with different orders of magnitude,
   and substitute them into Eq.(11). It can easily be shown that in case (A)
   \beq          
   S_{cl}(A)=4\pi me^{-2\phi_0}
   \eeq
   which is just the Bekenstein-Hawking entropy. However, in case (B),
   \beq            
   S_{cl}(B)=0
   \eeq
   which is just the result given by refs.[1,2]. Therefore, we have come to a conclusion that
   the two different results in fact come from taking limits in diffreent orders.
   Our results are in consistant with that given by Ghosh and Mitra recently[17].

   Unfortunately, we then get an entropy clash. In statistical physics and thermodynamics, entropy is
   a function of an equilibrium state only, and does not depend on the history or the
   process how the system arrives at the equilibrium state as well as the different
treatment of mathematics.
 From our discussions for
   the same EBH final state above, for different case (A) and (B) by taking limits
   in different orders, we find two different values for the entropy, namely the Bekenstein-Hawking
   entropy and zero. Of course, it would be most important to solve this clash.

   Notice that the above discussions are limited in the classical relativity.
     We recall that the Gibbs paradox of the entropy for a mixing ideal gas which
     had also appeared in classical statistics 
 and had been solved in quantum physics by considering the effects
     of identical particle. We hope to discuss the quantum entropy of the black hole
     and to see whether the quantum consideration will give some understanding of this clash.

     An early suggestion by 't Hooft[18] was that the fields propagating in the region
     just outside the horizon give the main contribution to the black hole entropy.
     The entropy arises from entanglement[19,20]. Many methods, for example, the brick wall
     model[18,5], Pauli-Villars regulator theory[6],etc., have been suggested to calculate
     the quantum effects of entropy in WKB approximation or in the one-loop approximation.
     Suppose the CDBH is enveloped by a scalar field, and the whole system, the hole and
     the scalar field, are filling in a cavity. The wave equation of the scalar field is
     \beq           
     \dpf{1}{\sqrt{-g}}\partial _\mu(\sqrt{-g}g^{\mu\nu}\partial _\nu\phi)-M^2\phi=0
     \eeq
     Substituting the metric Eq.(2) into Eq.(15), we find
     \beq          
     E^2(1-2me^{-\lambda r}+q^2e^{-2\lambda r})^{-1}f+\dpf{\partial}{\partial r}
      [(1-2me^{-\lambda r}+q^2e^{-2\lambda r})\dpf{\partial f}{\partial r}]-M^2f=0
      \eeq
      Introducing the brick wall boundary condition[18]
      \beqn
      \phi(x)=0\  {\rm at}\ r=r_+ +\ep\no     \\
      \phi(x)=0\  {\rm at}\ r=L  \no   \\  \no
      \eeqn
      and calculating the wave number $K(r,E)$ and the free energy $F$, we get
\beq                    
K^2(r,E)=(1-2me^{-\lambda r}+q^2e^{-2\lambda r})^{-1}[(1-2me^{-\lambda r}+q^2e^{-2\lambda r})^{-1}E^2-M^2]
\eeq
\beq             
F_{QM}=\dpf{\pi}{6\beta^2 \lambda}[\dpf{1}{2}\ln(R^2-2mR+q^2)+\dpf{m}{2\sqrt{m^2-q^2}}
	\ln\dpf{R-m-\sqrt{m^2-q^2}}{R-m+\sqrt{m^2-q^2}}]
\eeq
where $R=e^{\lambda(r_+ +\ep)}$, and $\ep\rightarrow 0$ is the coordinate cutoff
parameter.
To extend the above discussion to EBH, we are facing two limits $\epsilon\rightarrow 0$ and $q\rightarrow m$ again.
It can be proved that Eq(18) depends on the order of taking these two limits.
We find for case (A) which we take $\epsilon\rightarrow 0$ (i.e. $r\rightarrow r_+$) at first and the extremal limit afterwards,
\beq                 
F_{QM}(A)=-\dpf{\pi}{6\beta^2 \lambda}\ln(\dpf{1}{m\lambda \epsilon})
\eeq
but for case (B) which we adopt the extremal condition first and take $\epsilon\rightarrow 0$ afterwards
\beq              
F_{QM}(B)=-\dpf{\pi}{6\beta^2 \lambda}(\dpf{m}{m\lambda \epsilon}+\ln\dpf{1}{m\lambda \epsilon})
\eeq
Similar to the classical case, different expressions for free energy appear here due to different priority of taking different limits.
In order to compare with the other results, we replace the coordinate
variable $r$ by a proper variable $\rho$ through
${\rm d}\rho={\rm d}r/\sqrt{g(r)}$[18], we find that the proper cutoff $\ep'$
satisfies
\beq              
\ep=\dpf{\sqrt{m^2-q^2}\lambda\ep'^2}{2(m+\sqrt{m^2-q^2})}
\eeq
The linear divergence becomes quadratic.  Introducing
\beq          
\tilde{\ep}=\dpf{\sqrt{m^2-q^2}\lambda\ep'^2}{2}
\eeq
and substituting Eq.(22) into Eqs.(21),(20) and (19), we find Eqs(19)(20) can be reexpressed in the forms
\beq          
F_{QM}(A)=-\dpf{\pi}{6\beta^2\lambda}\ln\dpf{1}{\lambda\tilde{\ep}}
\eeq
\beq                  
F_{QM}(B)=-\dpf{\pi}{6\beta^2\lambda}(\dpf{m}{\tilde{\ep}}+\ln\dpf{1}{\lambda\tilde{\ep}})
\eeq
It is easy to find that the coordinate transformations do not change the different divergent behavior appeared in Eqs(19)(20) got from case (A) and case (B).
Here $\tilde{\ep}$ is also an infinitesimal quantity or $\tilde{\ep}\rightarrow 0$.
The basic difference is that $(\tilde{\ep})^{-1}$ includes not only the ultraviolet
divergence from the boundary condition but also the divergence from the extremal
limit. Through the entropy formula $S=\beta^2(\partial F/\partial\beta)$, we obtain
\beqn              
S_{QM}(A)=\dpf{\pi}{3\beta\lambda}\ln\dpf{1}{\lambda\tilde{\ep}}\\
S_{QM}(B)=\dpf{\pi}{3\beta\lambda}(\dpf{m}{\tilde{\ep}}+\ln\dpf{1}{\lambda\tilde{\ep}})
\eeqn
Comparing Eqs.(13)(14) with Eqs.(25)(26), we find that instead of a term being
proportional to the mass $m$ in the case (A) for the classical black hole [Eq.(13)],
we also find an additional term which is proportional to $m$ but linearly divergent for
$\tilde{\ep}$ in the case (B) with quantum correction for the black hole [Eq.(26)].
we adopt the viewpoint of the entanglement entropy [6,19,20] that the thermodynamical
entropy of the black hole system is in fact the sum $S=S_{cl}+S_{QM}$ so we
finally obtain
\beqn      
S(A)=S_{cl}(A)+S_{QM}(A)=4\pi e^{-2\phi_0}m+\dpf{\pi}{3\beta\lambda}\ln\dpf{1}{\lambda\tilde{\ep}}\\
S(B)=S_{cl}(B)+S_{QM}(B)=\dpf{\pi m}{3\beta\lambda\tilde{\ep}}+\dpf{\pi}{3\beta\lambda}
	\ln\dpf{1}{\lambda\tilde{\ep}}
\eeqn
Besides the same logarithmically divergent terms of $\tilde{\ep}$, the other two
terms in Eqs(27) and (28) are both proportional to the mass of the black hole. This result
is obviously in consistent with the results of refs.[5,22] in which they have
argued that $S=km$ from thermodynamics and statistical physics for EBH and $k$ is an
undetermined constant. However, as shown in Eq.(28), the additional term of the entropy
of scalar field includes the divergent factor $(1/\tilde{\ep})$.

As was pointed out earlier by Susskind and Uglum[23], the divergence of the quantum entropy
can be removed by renormalization of gravitational coupling $\tilde{G}$ for the
Schwarzchild black hole. Even though many authors have extended the argument of
$\tilde{G}$ renormalization to other NEBH[6,19], the renormalization for EBH has not
yet been realized. Note that the first term of the right hand side of Eq.(28) has
two factors, namely, $1/\beta$ and $\tilde{\ep}$, and the temperature $1/\beta$ of EBH
is zero. If one can prove that $1/\beta$ and $\tilde{\ep}$ have the same rate
to reach zero after some "renormalizations", this term will be finite and may compensate
the term $4\pi e^{-2\phi_0}m$ of the classical EBH because, in fact, the proportionality
coefficient $k$ of the entropy and the mass cannot be well determined for the
EBH[5]. Then, we get a possible explanation of the clash of the classical black hole in
the extremal limit. After taking the quantum effects into account, the entropy is a function of the
equilibrium state and is independent of the limits procedures as well as the processes through which the final
state is reached. A complete classical treatment of the entropy of the black hole
seems to be insufficient, and the quantum entropy must be involved. If we can find an expression of the entropy of
the whole black hole system through a suitable parameter which not only depends on $\epsilon$ and $q\rightarrow m$, but also on $\bar{h}$,
we speculate perhaps this clash can be resolved. Of course this is very difficult because we
have not yet got a successful quantum gravitational theory.

\vspace{1ex}
\hspace{0mm} This work was supported in part by NNSF of China.

\end{document}